\documentclass[twocolumn,english]{revtex4}
\usepackage[T1]{fontenc}
\usepackage[latin2]{inputenc}
\usepackage[letterpaper]{geometry}
\usepackage{textcomp}
\usepackage{amsmath}
\usepackage{amssymb}
\usepackage{graphicx}

\makeatletter

\providecommand{\tabularnewline}{\\}

\@ifundefined{textcolor}{}
{%
 \definecolor{BLACK}{gray}{0}
 \definecolor{WHITE}{gray}{1}
 \definecolor{RED}{rgb}{1,0,0}
 \definecolor{GREEN}{rgb}{0,1,0}
 \definecolor{BLUE}{rgb}{0,0,1}
 \definecolor{CYAN}{cmyk}{1,0,0,0}
 \definecolor{MAGENTA}{cmyk}{0,1,0,0}
 \definecolor{YELLOW}{cmyk}{0,0,1,0}
}

\makeatother

\usepackage{babel}
\begin{document}

\title{{\LARGE Nuclear wave packet quantum interference in the intense laser
dissociation of the $\mathrm{D}_{2}^{+}$ molecule}}

\author{Gábor J. Halász$^{1}$, Ágnes Vibók$^{2}$, }

\address{{\large $^{1}$}Department of Information Technology, University
of Debrecen, H-4010 Debrecen, PO Box 12, Hungary}

\address{{\large $^{2}$}Department of Theoretical Physics, University of
Debrecen, H-40410 Debrecen, PO Box 5, Hungary}

\author{Nimrod Moiseyev$^{3}$,}

\address{{\large $^{3}$}Schulich Faculty of Chemistry, Faculty of Physics,
and Minerva Center of Nonlinear Physics in Complex Systems, Technion
- Israel Institute of Technology, Haifa 32000, Israel}

\author{Lorenz S. Cederbaum$^{4}$}

\address{{\large $^{4}$}Theoretische Chemie, Physikalish-Chemisches Institut,
Universität Heidelberg, H-69120, Germany}
\begin{abstract}
Recently it has been recognized that electronic conical intersections
in molecular systems can be induced by laser light even in diatomics
\cite{nimrod1,milan1,gabi1}. As is known a direct consequence of
these accidental degeneracies is the appearence of nonadiabatic effects
which has a strong impact on the nuclear quantum dynamics. Studying
the photodissociation process of the $\mathrm{D}_{2}^{+}$ molecule,
we report here some novel and observable quantum interference phenomena
that arise from the topological singularity induced by a strong laser
field. 
\end{abstract}
\maketitle
Conical intersections (CI) between electronic potential energy surfaces
play a key mechanistic role in various basic molecular processes \cite{Horst1,Graham,Domcke1,Baer1}.
In this case the electronic states are coupled by the nuclear motion,
and the energy exchange between the fast electrons and the slow nuclei
becomes significant. Owing to the strong nonadiabatic couplings in
the close vicinity of these CIs the Born-Oppenheimer adiabatic approximation
breaks down. It was pointed out by Longuet-Higgins and Herzberg \cite{Longuet,Herzberg}
that each real adiabatic electronic state changes sign when transported
continuously along a closed path encircling the point of CI. Mead
and Truhlar associated this topological phase effect with the single
electronic state problem \cite{Mead} and Berry generalized the theory
\cite{Berry} , hence the name Berry phase. Ensuring that the electronic
wave function remains single valued one has to multiply it by a phase
factor and, as a consequence of it, this new electronic eigenfunction,
instead of being real, becomes complex. The fact that the electronic
eigenfunctions are changed has a direct effect on the nuclear dynamics
even if it takes place in a single potential energy surface. Therefore,
the presence of the Berry phase in a molecular system can be considered
as a clear fingerprint of the CI. 

CIs in molecules can be formed only if the studied molecular system
possesses at least two independent nuclear degrees of freedom. Therefore,
for a diatomic molecule that has only one nuclear vibrational coordinate,
it is not possible for two electronic states of the same symmetry
to become degenerate (unless they belong to a doubly degenerate irreducible
representation (e.g.$\pi$). However, this latter statement is true
only in a field free space. The rotation of diatomic molecules exposed
to strong laser fields can serve as an additional degree of freedom
because the interaction of the induced or permanent dipole moment
of the system with the electric field leads to an effective torque
toward the polarization direction. It was presented in earlier works
\cite{nimrod1,milan1} that CIs can be formed both by running or standing
lasers waves even in diatomics. In these situations additional to
the nuclear vibration the rotational motion or the position of the
center of mass provide the missing degree of freedom to make possible
the formation of a CI. This process is easily understood within the
framework of the dressed state representation. Describing the molecule-light
interaction in this model, i.e., when the light field is explicitly
included into the Hamiltonian, the change of nuclear dynamics due
to the light field can be considered as arising from the appearance
of a ``light-induced conical intersection'' (LICI). The energetic
and spatial position of these light-induced CIs can be controlled
by varying the parameter settings of the laser field. Detailed theoretical
investigations demonstrate that these light-induced CIs have a strong
impact on the nuclear dynamics \cite{gabi1,gabi2,gabi3,gabi4,gabi5}. 

\begin{figure}
\begin{centering}
\includegraphics[width=0.48\textwidth]{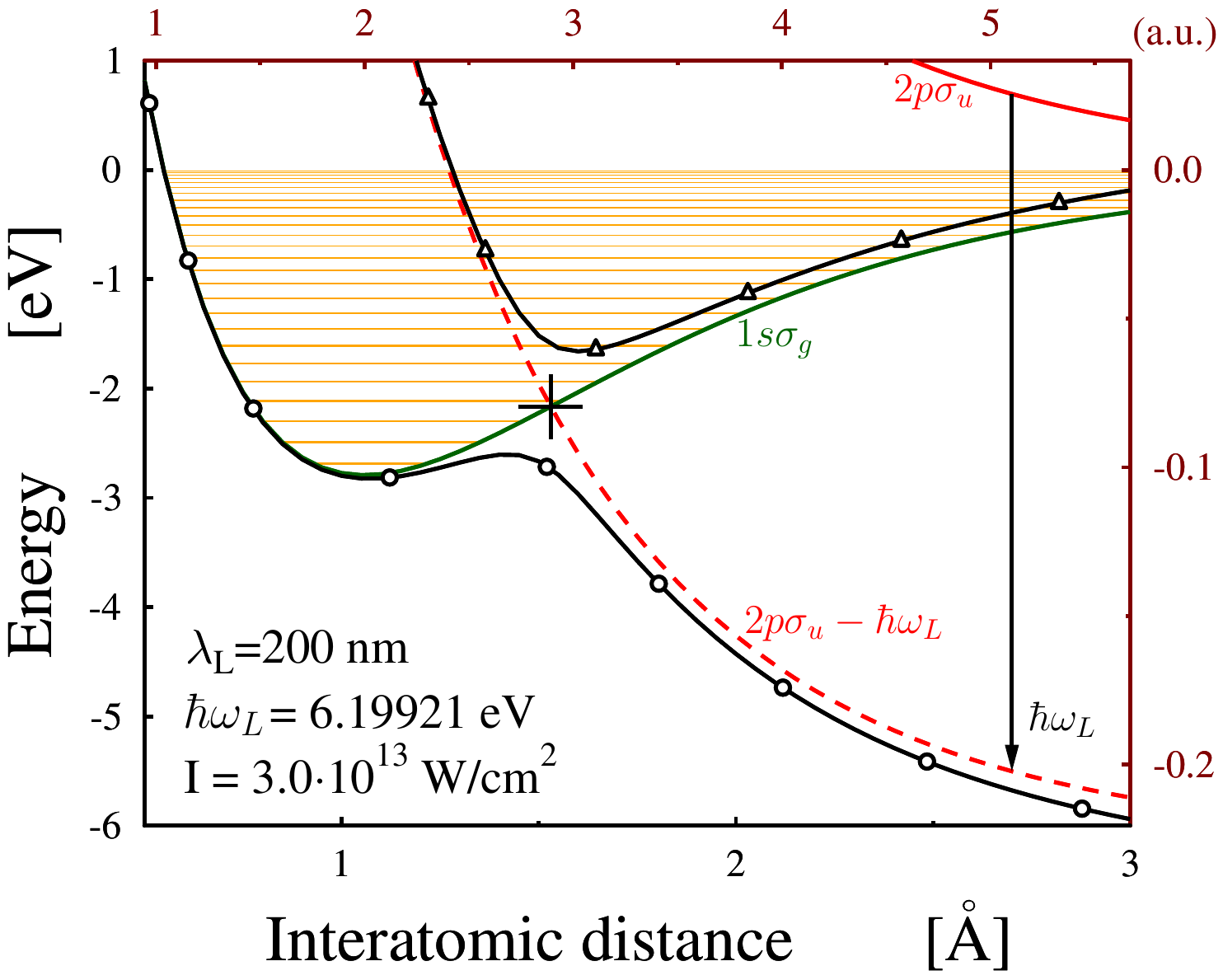}
\par\end{centering}

\caption{\label{fig:1}Potential energies of the $\mathrm{D}_{2}^{+}$ molecule
as a function of interatomic separation. Diabatic energies of the
ground $\left(1s\sigma_{g}\right)$ and the first excited $\left(2p\sigma_{u}\right)$
states are displayed with solid green and red lines, respectively.
The field dressed excited state ($2p\sigma_{u}-\hbar\omega_{L}$ ;
dashed red line) forms a light induced conical intersection (LICI)
with the ground one. For the case of field intensity of $3\times10^{13}\frac{W}{cm^{2}}$
a cut of the adiabatic surfaces at $\theta=0$ (parallel with the
field) are also shown by solid black lines marked with circles (lower)
and triangles (upper). We denoted with a cross the position of the
LICI ($R_{LICI}=1.53\textrm{\AA}=2.891a.u.$ and $E_{LICI}=-2.16611eV$).}
\end{figure}

In our former works we have pointed out that the appearence of the
topological or Berry phase in a molecular system is a clear signature
of the CI independently of whether it is a natural or a laser-induced
one \cite{gabi1,gabi2}. Applying the line integral technique \cite{Baer2}
in the Floquet representation we calculated the Berry phase in diatomics
for a LICI situation. One of the clearest observable manifestation
of the topological or Berry phase in a molecular system is the appearence
of a quantum interference effect \cite{Chiao1,Chiao2}. In order to
illustrate this fundamental effect for the present LICI situation,
we have chosen as an explicit example the photodissotiation of the
D$_{2}^{+}$molecule, which has been extensively studied for more
than two decades \cite{Bandrauk,Sandig,atabek,Esry2,Esry1,Fischer}.
In addition, it has also an advantage: due to its relative simplicity
one can study the new light-induced nonadiabatic phenomena separately
from other processes. In our model calculation we will consider the
two relevant electronic states of the $\mathrm{D}_{2}^{+}$ ion, which
are characterized by a ground ($V_{1}=1s\sigma_{g}$) and a first
excited ($V_{2}=2p\sigma_{u}$) potential energy curves, respectively.
For studying the dissociation process the following mechanism will
be considered (Fig. \ref{fig:1}): initially the $\mathrm{D}_{2}^{+}$
ion is in its ground electronic ($1s\sigma_{g}$) and rotational state
combined with one of its vibrational eigenstates. Then, it is excited
from this $1s\sigma_{g}$ state by resonant laser field to the repulsive
$2p\sigma_{u}$ state. As a result of this process the $1s\sigma_{g}$
and $2p\sigma_{u}$ electronic states are resonantly coupled by the
laser. The non-vanishing dipole matrix elements are responsible for
the light-induced electronic transitions. In the space of these two
electronic states the following time-dependent Hamiltonian holds for
the rovibronic nuclear motions:

\begin{align}
H & =\left(\begin{array}{cc}
-\frac{1}{2\mu}\frac{\partial^{2}}{\partial R^{2}}+\frac{L_{\theta\varphi}^{2}}{2\mu R^{2}} & \;0\\
0 & \;-\frac{1}{2\mu}\frac{\partial^{2}}{\partial R^{2}}+\frac{L_{\theta\varphi}^{2}}{2\mu R^{2}}
\end{array}\right)+\label{eq:Hamilton}\\
 & \left(\begin{array}{cc}
V_{1}(R) & -\epsilon_{0}f(t)d(R)\cos\theta\cos\omega_{L}t\\
-\epsilon_{0}f(t)d(R)\cos\theta\cos\omega_{L}t & V_{2}(R)
\end{array}\right).\nonumber 
\end{align}
Here, R and ($\theta,\varphi$) are the molecular vibrational and
rotational coordinates, respectively, $\mu$ is the reduced mass,
and $L_{\theta\varphi}$ denotes the angular momentum operator of
the nuclei. Here $\theta$ denotes the angle between the polarization
direction and the direction of the transition dipole and thus one
of the angles of rotation of the molecule. $V_{1}(R)$ ($1s\sigma_{g}$)
and $V_{2}(R)$ ($2p\sigma_{u}$) are the two electronic states coupled
by the laser (whose frequency is $\omega_{L}$ and amplitude is $\epsilon_{0}$),
$f(t)$ is the envelop function and $d(R)$$\left(=-\left\langle \psi_{1}^{e}\left|\sum_{j}r_{j}\right|\psi_{2}^{e}\right\rangle \right)$
is the transition dipole matrix element ($e=m_{e}=\hbar=1;$ atomic
units are used throughout the article). The potential energies $V_{1}(R)$
and $V_{2}(R)$ and the transition dipole moment were taken from \cite{dipol,pot}.
The dressed state representation is very illustrative and helps to
understand the essence of the light-induced nonadiabatic effects.
In this representation the laser light shifts the energy of the $2p\sigma_{u}$
repulsive excited potential curve by $\hbar\omega_{L}$ and a crossing
between the diabatic ground and the diabatic shifted excited potential
energy curves is formed. After diagonalizing the potential energy
matrix the adiabatic potential surfaces $V_{lower}$ and $V_{upper}$
can be obtained (Fig. \ref{fig:1}). 
\begin{table}
\centering{}\caption{\label{tab:1}Energies of the studied vibrational levels of the $\mathrm{D}_{2}^{+}$
molecule (in $eV$). }
\begin{tabular}{|l||c|c|c|c|c|}
\cline{2-6} 
\multicolumn{1}{l||}{} & $\nu=0$ & $\nu=1$ & $\nu=2$ & $\nu=3$ & $\nu=4$\tabularnewline
\hline 
\hline 
$E_{\nu}$  & -2.6900 & -2.4905 & -2.2985 & -2.1145 & -1.9380\tabularnewline
\hline 
\end{tabular}
\end{table}
These two surfaces can cross each other, giving rise to a conical
intersection whenever the conditions $\cos\theta=0$, $(\theta=\pi/2)$
and $V_{1}(R)=V_{2}(R)-\hbar\omega_{L})$ are simultaneously fulfilled. 

The characteristic features of the LICI can be changed by varying
the frequency and intensity of the laser field. Increasing the frequency,
for example, moves the CI to a smaller internuclear distance and to
a smaller energetic position. The steepness of the CI cone formed
by the adiabatic surfaces, which is related to the strength of the
nonadiabatic coupling, can be controlled by the laser intensity. To
demonstrate the impact of the LICI on the photodissociation dynamics
one has to solve the time-dependent nuclear Schrödinger equation (TDSE)
with the Hamiltonian $\hat{H}$ given by Eq. (\ref{eq:Hamilton}).
One of the most efficient approaches for solving the time-dependent
nuclear Schrödinger equation is the MCTDH (multi configuration time-dependent
Hartree) method \cite{dieter1,dieter2,dieter3,dieter4}. To describe
the vibrational degree of freedom we have used FFT-DVR (Fast Fourier
Transformation-Discrete Variable Representation) with $N_{R}$ basis
elements distributed within the range from 0.1 a.u. to 80 a.u. for
the internuclear separation. The rotational degree of freedom was
described by Legendre polynomials $\left\{ P_{J}(\cos\theta)\right\} _{j=0,1,2,\cdots,N_{\theta}}$.
These so called primitive basis sets ($\chi$) were used to represent
the single particle functions ($\phi$), which in turn were used to
represent the wave function:
\begin{eqnarray}
\phi_{j_{q}}^{(q)}(q,t) & = & \sum_{l=1}^{N_{q}}c_{j_{q}l}^{(q)}(t)\;\chi_{l}^{(q)}(q)\qquad q=R,\,\theta\label{eq:MCTDH-wf}\\
\psi(R,\theta,t) & = & \sum_{j_{R}=1}^{n_{R}}\sum_{j_{\theta}=1}^{n_{\theta}}A_{j_{R},j_{\theta}}(t)\phi_{j_{R}}^{(R)}(R,t)\phi_{j_{\theta}}^{(\theta)}(\theta,t).\nonumber 
\end{eqnarray}
In our numerical calculations we have used $N_{R}=2048$ and, depending
on the field intensity, $N_{\theta}=6,\cdots,70$. On both diabatic
surfaces and for both degrees of freedom a set of $n_{R}=n_{\theta}=3,\cdots,25$
single particle functions were used to build up the nuclear wave function
of the system. (The actual value of $N_{\theta}$ and $n_{R}=n_{\theta}$
was chosen depending on the peak field intensity $I_{0}$.) All the
calculations were correctly converged with these parameters. 

\begin{figure*}
{\Large (a)}\hspace{-1cm}\includegraphics[width=0.48\textwidth]{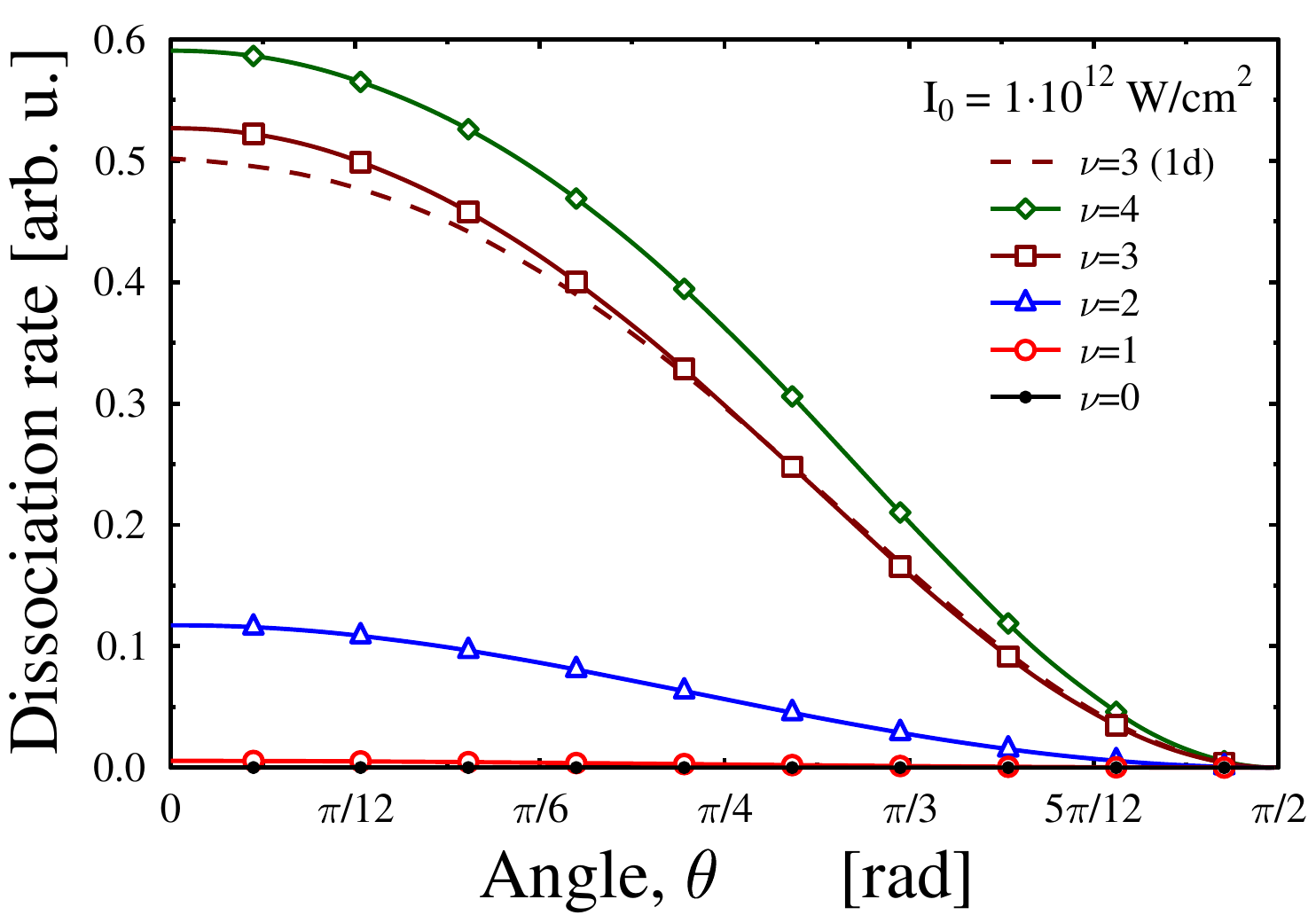}\hfill{}{\Large (b)}\hspace{-1cm}\includegraphics[width=0.48\textwidth]{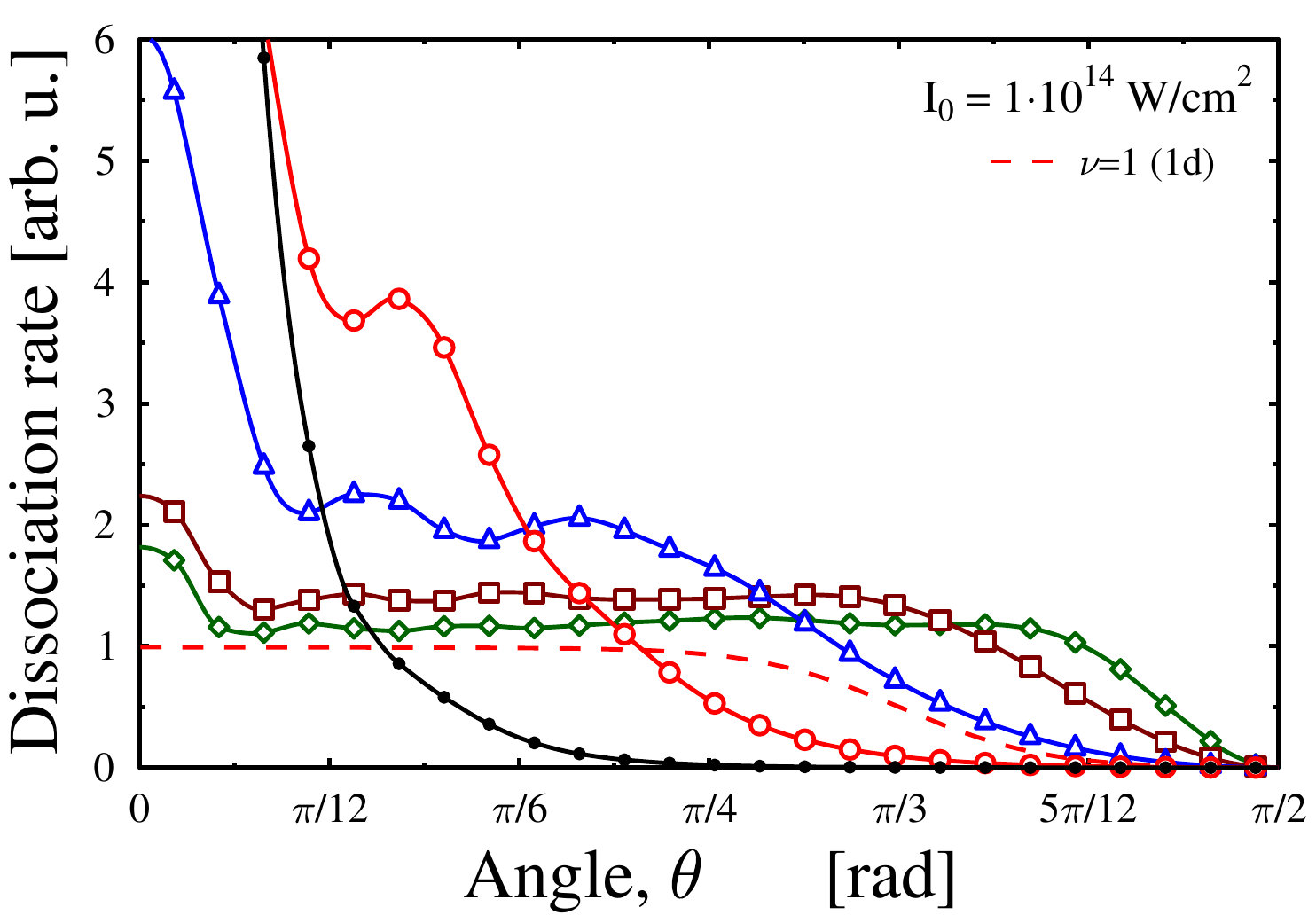}

\caption{\label{fig:2}Fragment angular distributions of the dissociating $\mathrm{D}_{2}^{+}$
molecule for two different intensities ($1\times10^{12}$ and $1\times10^{14}\, W/cm^{2}$)
and five different vibrational states ($\nu=0,1,2,3,4$). One dimensional
(1d) calculations are presented for $\nu=3$ at $1\times10^{12}$
and $\nu=1$ at $1\times10^{14}\, W/cm^{2}$ intensities. }
\end{figure*}
The solution of equation (\ref{eq:MCTDH-wf}) is used to calculate
the total dissociation probability and the angular distribution of
the photofragments \cite{dieter3}: 

\begin{equation}
P_{diss}=\intop_{0}^{\infty}dt<\psi(t)|W|\psi(t)>\label{eq:ker}
\end{equation}
where $-iW$ is the complex absorbing potential (CAP) applied at the
last 5 a.u. of the grid related to the vibrational degree of freedom,
and

\begin{equation}
P(\theta_{j})=\frac{1}{w_{j}}\intop_{0}^{\infty}dt<\psi(t)|W_{\theta_{j}}|\psi(t)>\label{eq:angdist}
\end{equation}
where $-iW_{\theta_{j}}$ is the projection of the CAP to a specific
point of the angular grid $\left(j=0,..N_{\theta}\right)$, and $w_{j}$
is the weight related to this grid point according to the applied
DVR.

\begin{table}
\caption{\label{tab:2}Dissociation probability as a function of the field
intensity. (Maximum intensities are given in $W/cm^{2}$.)}

\centering{}%
\begin{tabular}{|c||c|c|c|c|c|}
\hline 
$I_{0}$ & $\nu=0$ & $\nu=1$ & $\nu=2$ & $\nu=3$ & $\nu=4$\tabularnewline
\hline 
\hline 
$1\cdot10^{11}$ & $1.9\cdot10^{-6}$ & $1.7\cdot10^{-4}$ & $0.0037$ & $0.0232$ & $0.0291$\tabularnewline
\hline 
$3\cdot10^{11}$ & $5.6\cdot10^{-6}$ & $5.1\cdot10^{-4}$ & $0.0113$ & $0.0669$ & $0.0830$\tabularnewline
\hline 
$1\cdot10^{12}$ & $1.9\cdot10^{-5}$ & $0.0018$ & $0.0379$ & $0.1948$ & $0.2339$\tabularnewline
\hline 
$3\cdot10^{12}$ & $6.6\cdot10^{-5}$ & $0.0060$ & $0.1150$ & $0.4250$ & $0.4715$\tabularnewline
\hline 
$1\cdot10^{13}$ & $3.4\cdot10^{-4}$ & $0.0299$ & $0.3612$ & $0.7143$ & $0.7105$\tabularnewline
\hline 
$3\cdot10^{13}$ & $0.0042$ & $0.2081$ & $0.7237$ & $0.9009$ & $0.8757$\tabularnewline
\hline 
$1\cdot10^{14}$ & $0.3051$ & $0.7873$ & $0.9502$ & $0.9868$ & $0.9794$\tabularnewline
\hline 
$3\cdot10^{14}$ & $0.9251$ & $0.9781$ & $0.9962$ & $0.9993$ & $0.9956$\tabularnewline
\hline 
\end{tabular}
\end{table}
In all of our calculations the initial nuclear wave packet (at $t=0\, fs$)
was chosen to be in its rotational ground state ($J=0$) and in some
of its vibrational eigenstate $\left(\nu=0,1,2,3,4\right)$. In this
way the angular distribution of the photofragments can provide an
accurate information about the dissociation of single vibrational
levels \cite{Sandig}. Nonaligned (isotropic) molecules and linearly
polarized Gaussian laser pulses centered around $t=0\, fs$ were used
in the numerical simulations. The center wavelength and the pulse
duration at full width of half maximum (FWHM) are 200 nm and $t_{pulse}=30$,
respectively. We note here that for these values of the laser parameters
the $\nu=0,1,2$ vibrational levels are below the energy of the LICI
$\left(E_{LICI}=-2.1661eV\right)$ and the $\nu=3,4$ states are above
of it (see Fig. \ref{fig:1} and Table \ref{tab:1}). We calculated
the total dissociation probability $P_{diss}$ of the photofragments.
Results for different intensities are presented in Table \ref{tab:2}.
At low intensities the dissociation probabilities of the low vibrational
levels $\left(\nu=0,1\right)$ are small due to the potential barrier.
In contrast, the dissociation rate of the $\nu=3$ and $\nu=4$ vibrational
levels are relatively large for intensity values\textbf{\LARGE{} }greater
than $1\times10^{12}\frac{W}{cm^{2}}$. This feature can easily be
explained borrowing the light induced potential (LIP) picture \cite{Bandrauk}.
The molecules are exposed to small and intermediate light intensities
and follow the diabatic dissociation path while for high enough intensities
they rather tend to follow the adiabatic path. For small light intensities
the dissociation behaves linearly with the intensity, whereas at higher
intensities up to $10^{14}W/cm^{2}$ this quantity is strongly nonlinear.
In the present case the $\nu=0,1,2$ levels lie in the lower adiabatic
potential. As for the other states, they are in the gap between the
two adiabatic potential curves. The number of vibrational states in
the gap region depends on the value of the intensity. Increasing the
intensity, the gap between the two adiabatic potential curves is increasing.
As a consequence, the so-called bond softening mechanism occurs for
states $\nu=0,1,2$, (see in Table \ref{tab:2} for $10^{14}W/cm^{2}$),
while for higher vibrational states, which are not in the gap region,
the well known trapping effect or bond hardening emerges. For those
vibrational states that are in the gap region the transition probability
is relatively large. Neither bond hardening nor bond softening occurs.
For high enough intensities almost all molecules from these states
dissociate following the nonadiabatic path. In our case the $\nu=3$
and $\nu=4$ vibrational states are in the gap region and provide
almost 100\% dissociation rate at $10^{14}W/cm^{2}$. As a next step
we calculate the fragment angular distribution. The obtained results
are summarized in Fig. \ref{fig:2}. Regarding the structures of these
curves the following observation can be made: Curves are very smooth
for $1\times10^{12}W/cm^{2}$ , but for large enough intensities $\left(1\times10^{14}W/cm^{2}\right)$
some distinct modulations appear on them. Similar results were obtained
in a very recent paper of Fischer and coworkers \cite{Fischer}. They
have studied the photodissociation of the $\mathrm{H}_{2}^{+}$ molecule
solving the TDSE including the rotational dynamics. They considered
vibrationally and rotationally excited initial wave packet, $\left(\nu=4,\,\, J=2\right)$
and the results obtained are presented in Fig. 3c of \cite{Fischer}).
Their conditions are very similar to ours, but since the applied wavelength
is larger $\left(\lambda=266nm\right)$ the position of the LICI is
shifted to the right. Consequently, the $\nu=3$ level is still below
the energy of the CI but the $\nu=4$ state lies already above it.
After carefully checking the laser parameters they used, it seems
that the $\nu=4$ level of H$_{2}^{+}$ sits in the gap between the
two light induced adiabatic potentials. The $\nu=4$ level plays the
same role in their study as $\nu=3$ in our case. Both sit on the
gap region. Correspondingly, Fig. 3c in (\cite{Fischer}) is very
similar to ours (Fig. \ref{fig:2}, I=$1\times10^{14}W/cm^{2}$, curve
for $\nu=3$), despite of the fact that two different theoretical
methods were used for the solution of the TDSE. The distinct modulations
and the abrupt decreases for angles $\theta\geq60^{\circ}$ appear
in both pictures. There is only one difference between the two initial
wave packets. Our wave packet is in its rotationally ground state
$(J=0)$, whereas theirs is rotationally excited $(J=2)$. This, however
does not mean essential difference, because the LICI introduces an
intense nonadiabatic coupling via mixing the rotational and vibrational
motions on both electronic surfaces and, therefore, even if one starts
with $J=0$, afterwards it would have J values up to 30 or 40. To
understand more deeply the modulations on the curves, we have performed
the analysis of the nuclear density function $\left|\psi(R,\theta,t)\right|^{2}\left(=\left|\psi^{1s\sigma_{g}}(R,\theta,t)\right|^{2}+\left|\psi^{2p\sigma_{u}}(R,\theta,t)\right|^{2}\right)$.
\begin{figure*}
\begin{centering}
\begin{minipage}[c]{0.9\textwidth}%
\begin{center}
\includegraphics[width=0.32\textwidth]{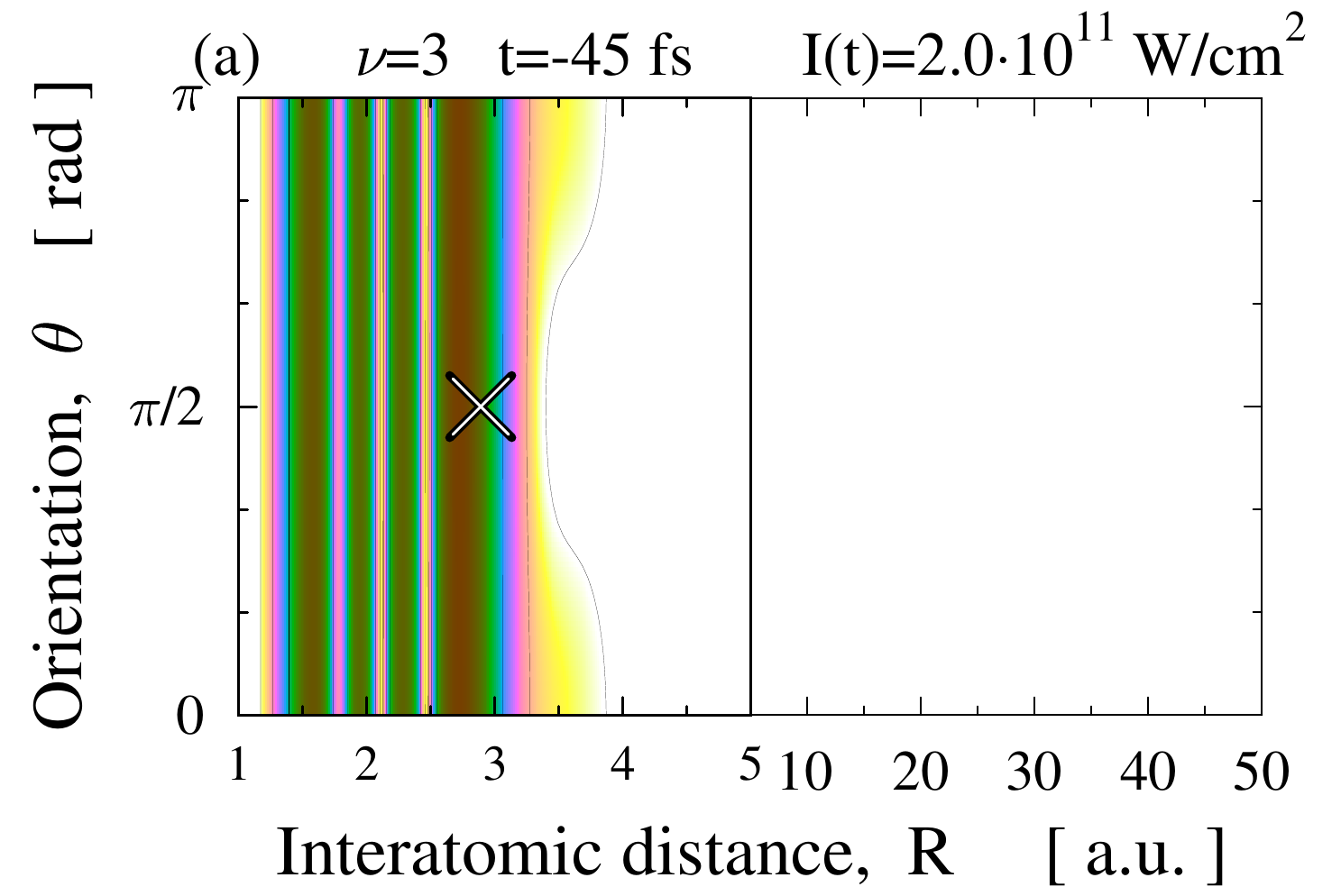}\hfill{}\includegraphics[width=0.32\textwidth]{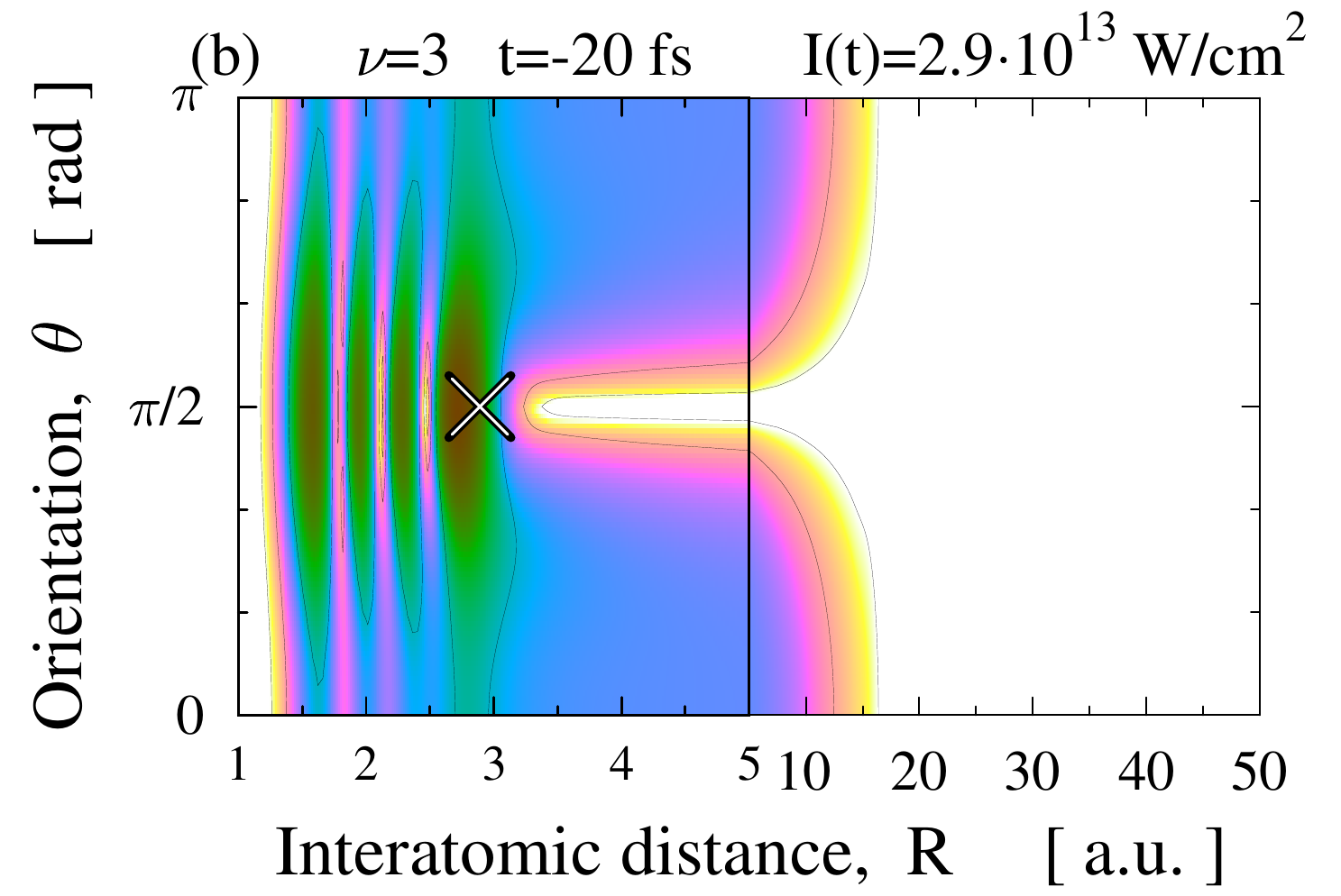}\hfill{}\includegraphics[width=0.32\textwidth]{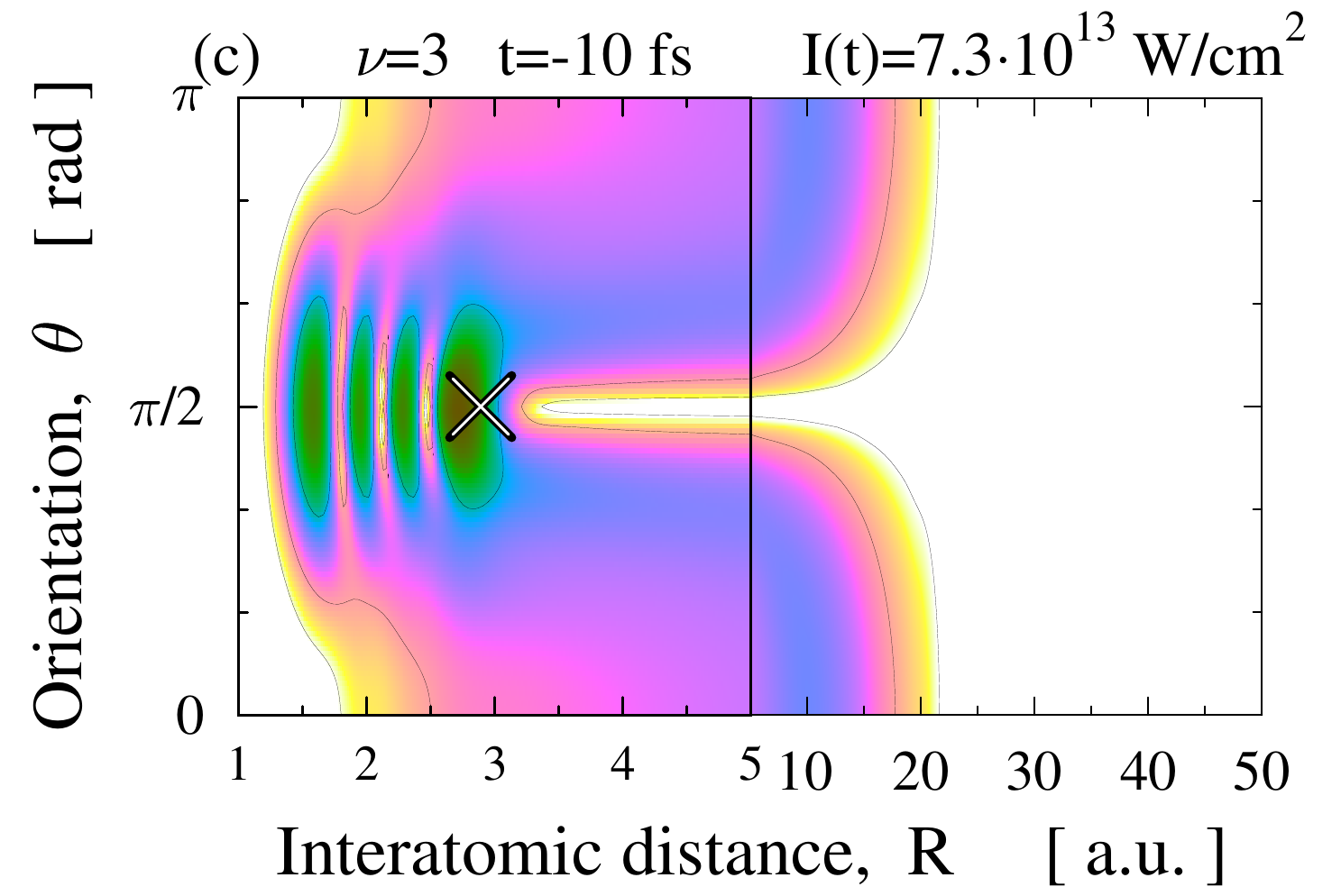}
\par\end{center}

\begin{center}
\includegraphics[width=0.32\textwidth]{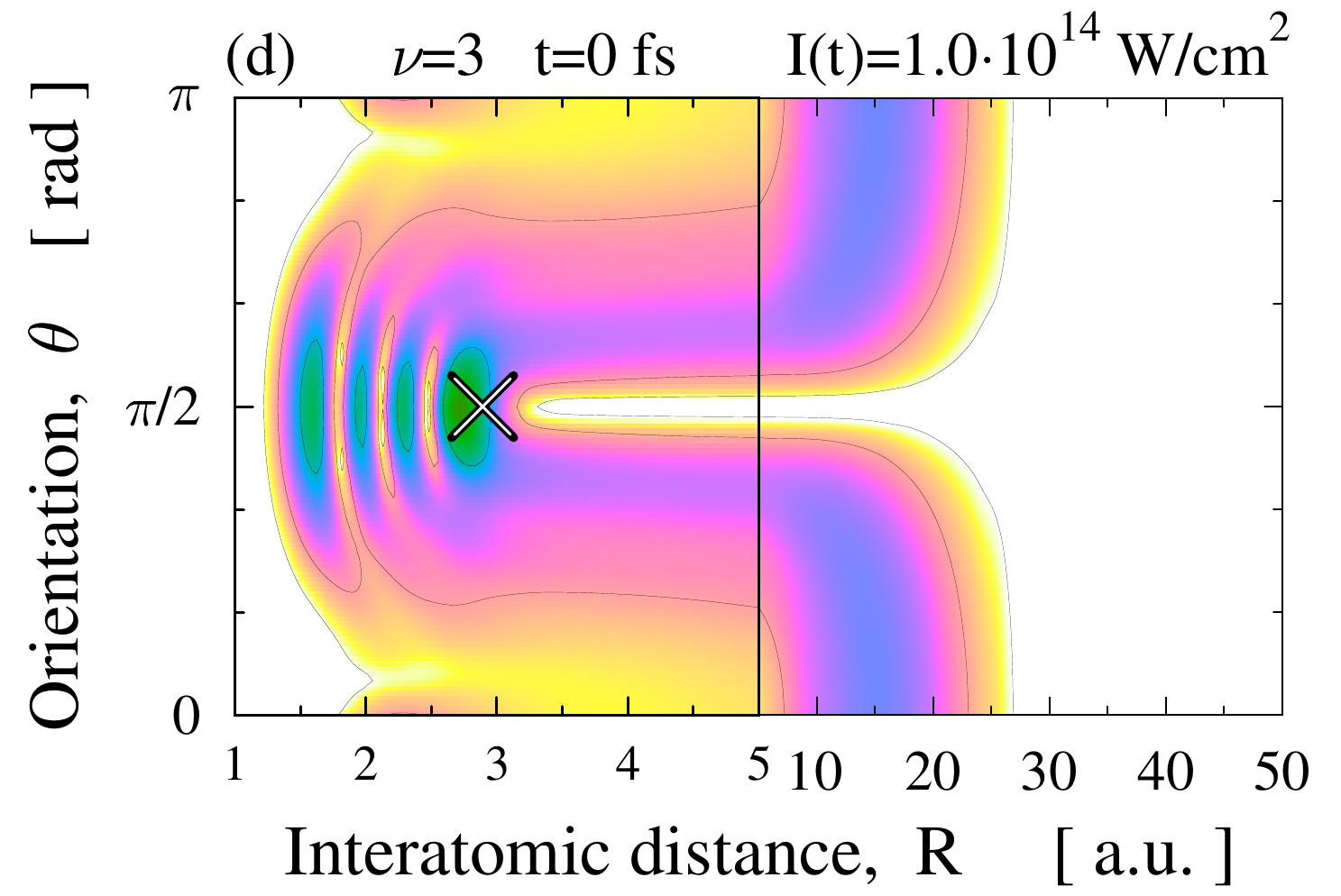}\hfill{}\includegraphics[width=0.32\textwidth]{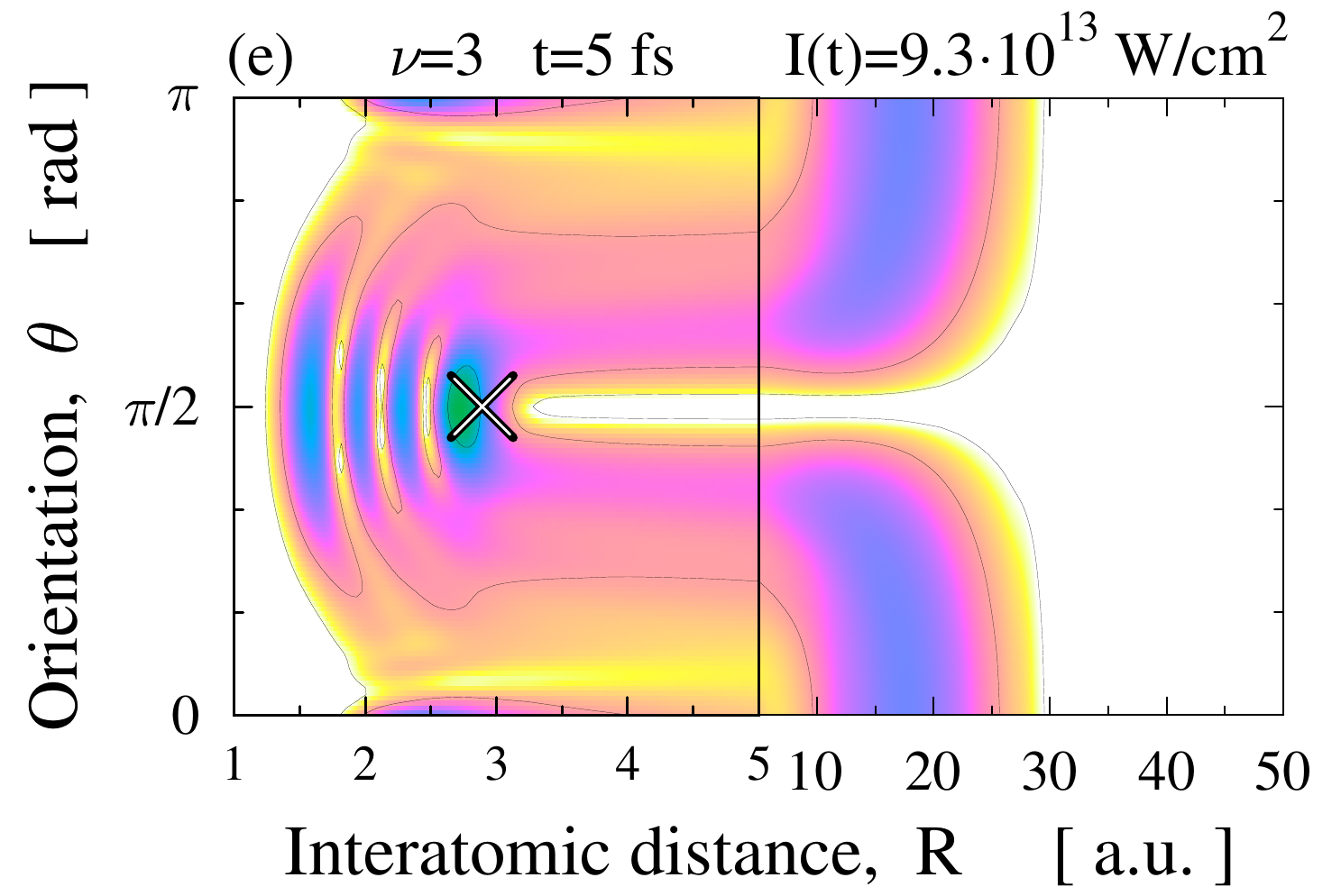}\hfill{}\includegraphics[width=0.32\textwidth]{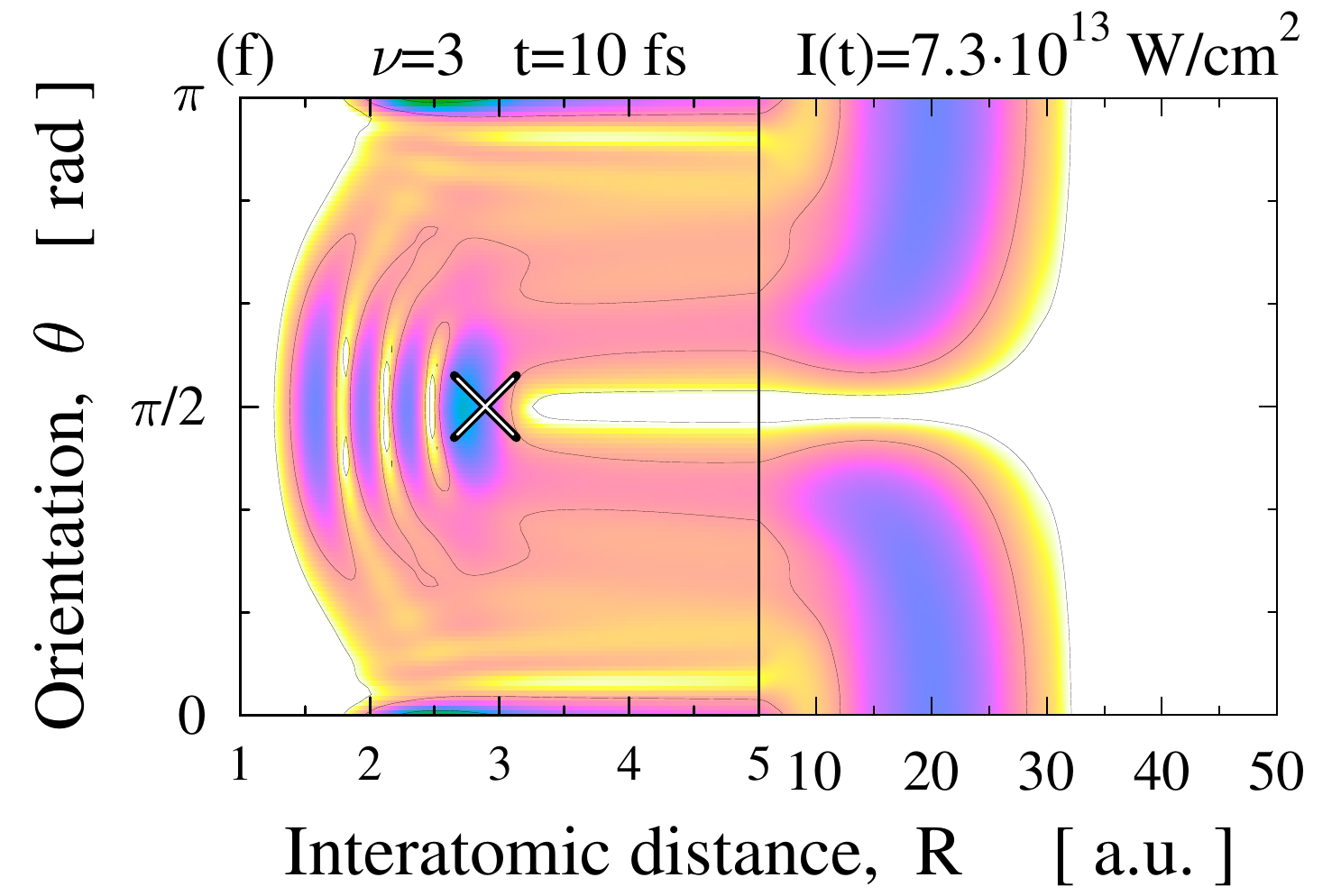}
\par\end{center}

\begin{center}
\includegraphics[width=0.32\textwidth]{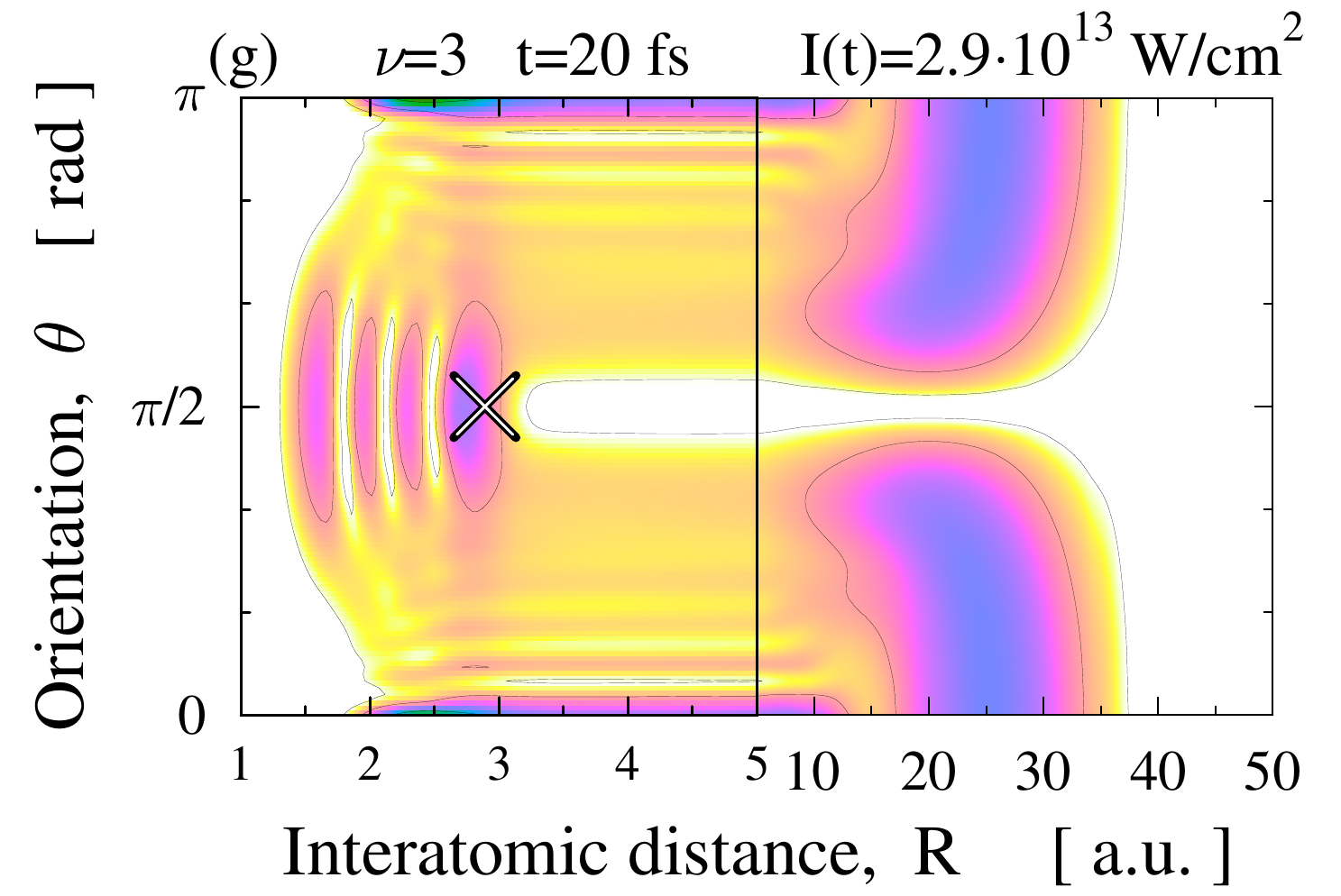}\hfill{}\includegraphics[width=0.32\textwidth]{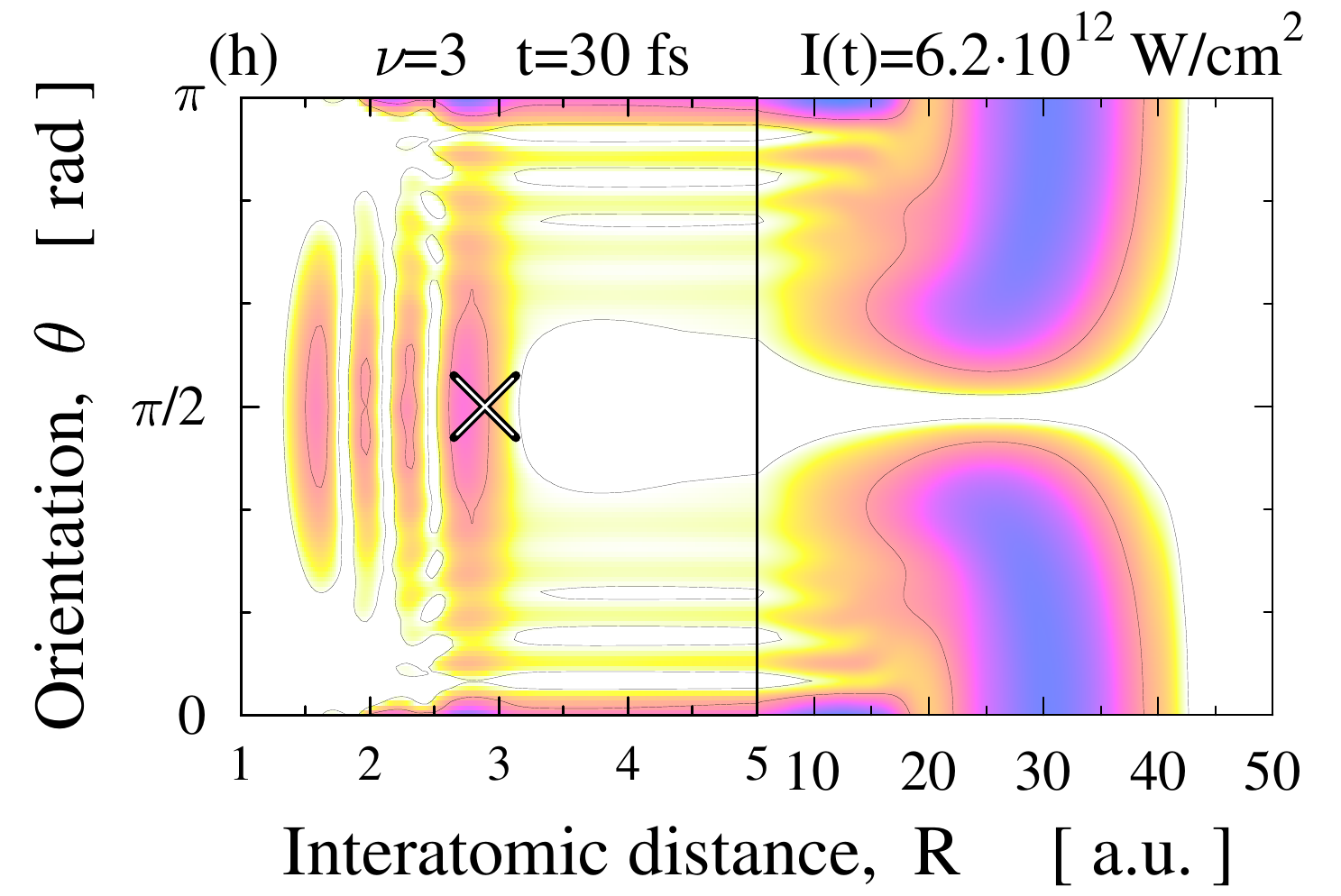}\hfill{}\includegraphics[width=0.32\textwidth]{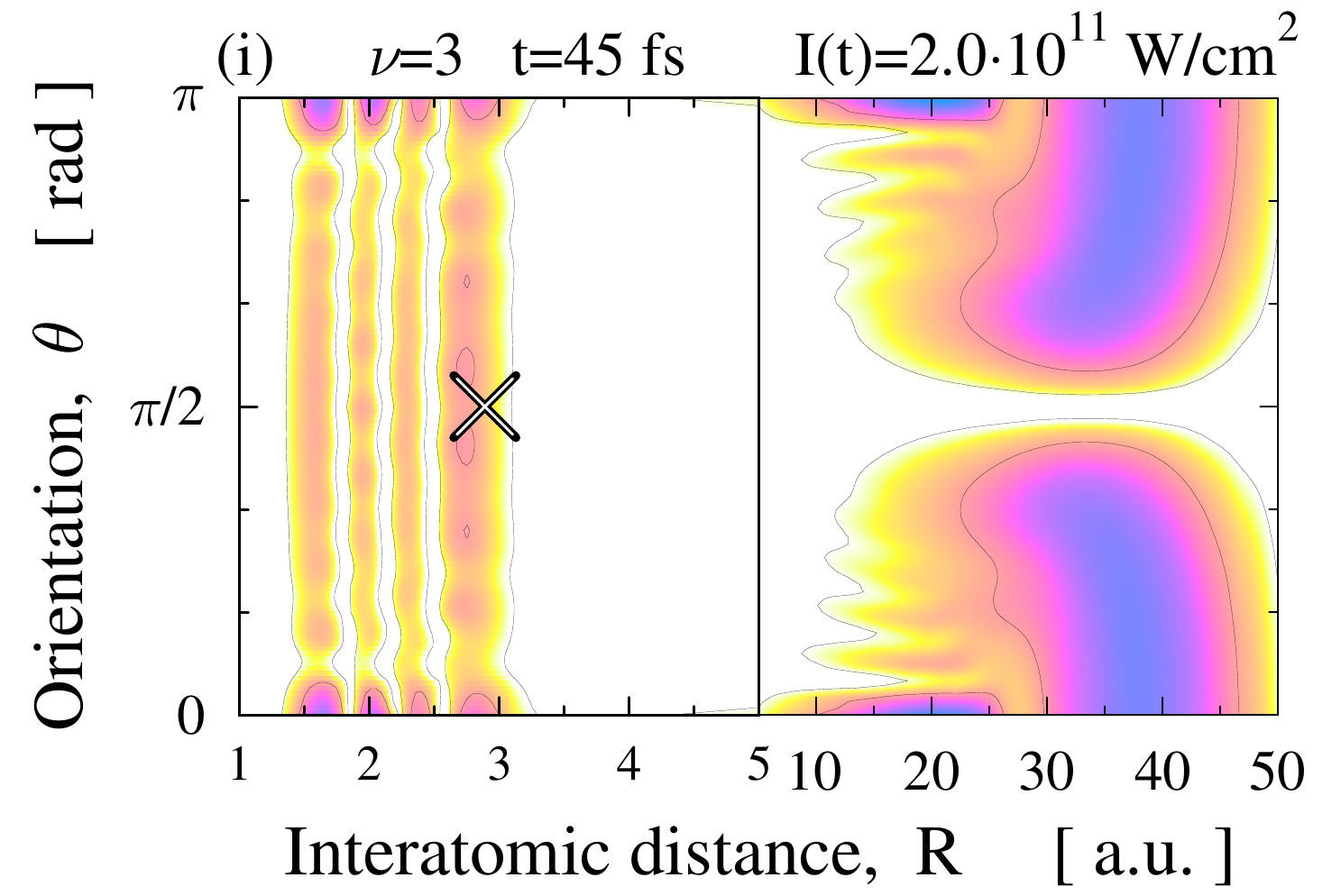}
\par\end{center}%
\end{minipage}\hfill{}%
\begin{minipage}[c]{0.09\textwidth}%
\begin{center}
\includegraphics[width=1\textwidth]{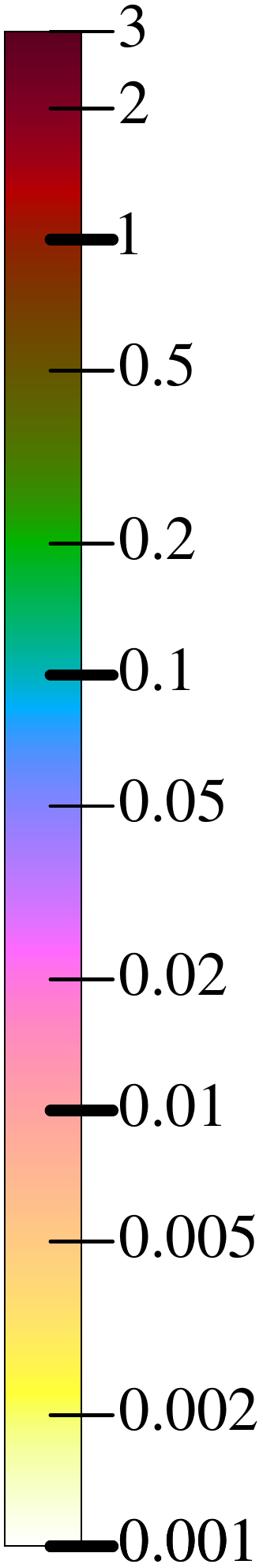}
\par\end{center}%
\end{minipage}
\par\end{centering}

\caption{\label{fig:3}Snapshots from the real-time evolution of the interference
appearing in the angular distribution of the dissociated particles.
I=$1\times10^{14}W/cm^{2}$ intensity and $\nu=3$ vibrational state
are applied. The cross denotes the position of the LICI. }
\end{figure*}
Obtained results are illustrated in Fig. \ref{fig:3} with snapshots
from the structure of the nuclear wave packet density $\left|\psi(R,\theta,t)\right|^{2}$
at different times. It is seen that the nuclear wave packet passes
through the dissociation region, but owing to the strong nonadiabatic
topological effect of the LICI, it is unable to move unhindered. It
splits up in front of the LICI region, bypasses both sides of the
LICI and then its two components exhibit a complex motion. On the
one hand they move horizontally from left to the right, which would
represent the pure dissociation, but on the other hand, they also
move in vertical direction from top to bottom and vice versa. Due
to the rotation induced by the external electric field rotational
nodes being formed roughly after the maxima of the intensity is reached
($t=0\, fs$). It means that at given directions the wave function
disappears from the bounded region of the configuration space. As
a consequence of it, the dissociation in these directions will also
be reduced. But in contrast, above and below of these nodes the value
of the wave packet density is significantly different from zero. So-called
quantum interference effects appear, maxima turn into minima and vice
versa. These quantum interference effects are the sources of the bumpy
structure appearing in the angular distribution of the photofragments
(see Fig. \ref{fig:2}b). We note here, that for symmetry reasons
while the one photon processes are dominant in the laser-matter interaction
the nuclear wave packet should vanish at $\theta=\pi/2$ in the dissociation
region. 

\begin{figure*}[t]
\begin{centering}
\begin{minipage}[c]{0.92\textwidth}%
\begin{center}
\includegraphics[width=0.32\textwidth]{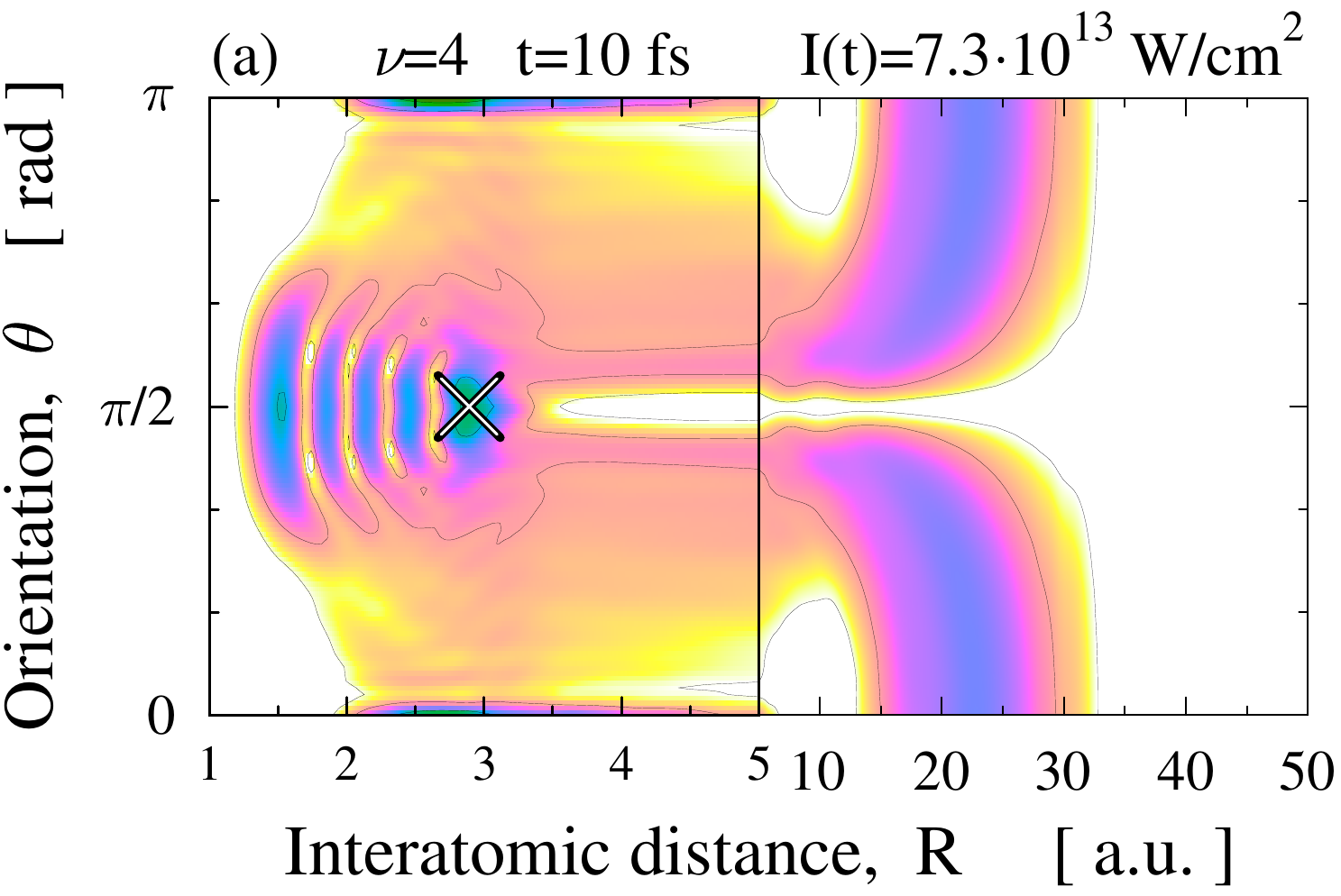}\hfill{}\includegraphics[width=0.32\textwidth]{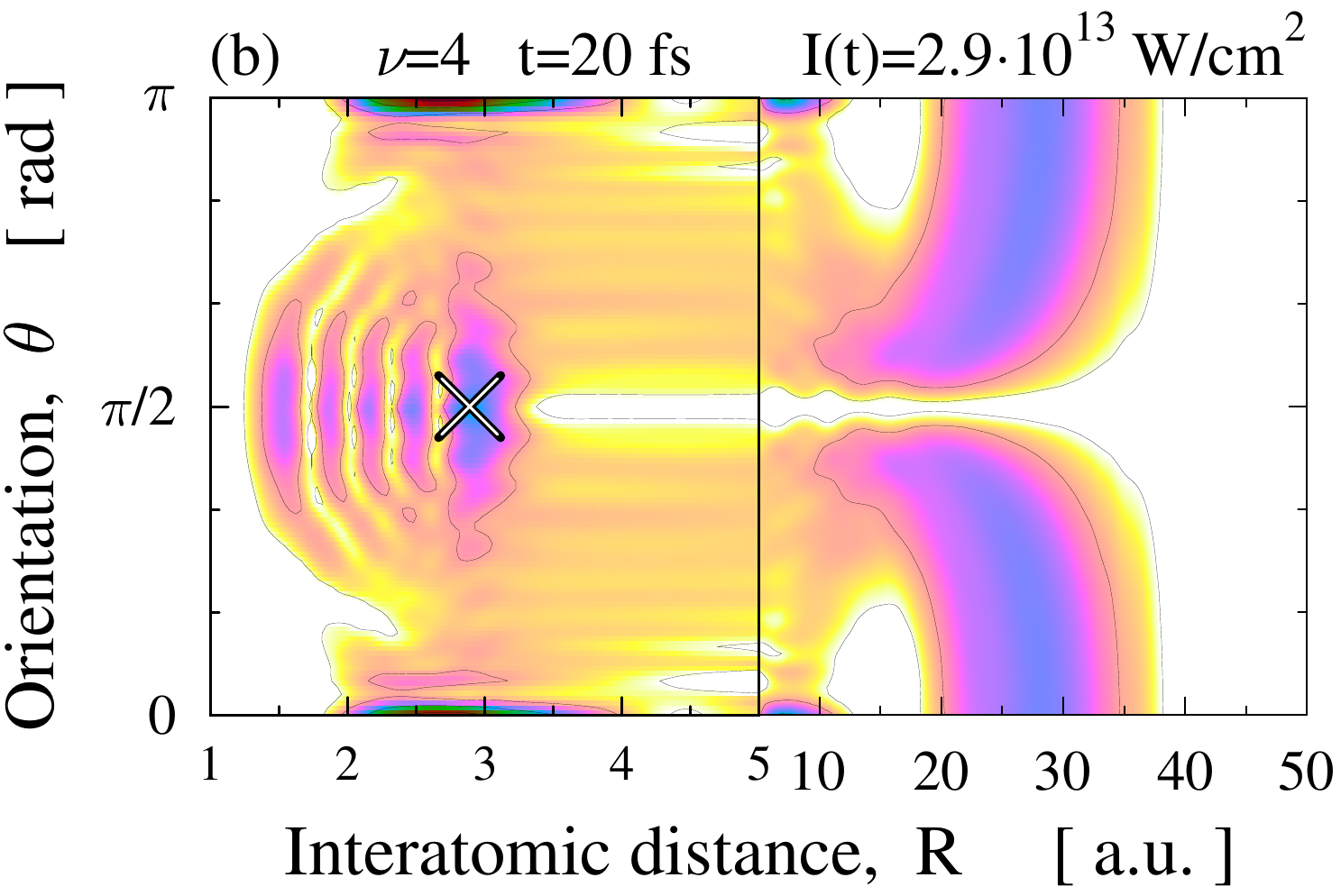}\hfill{}\includegraphics[width=0.32\textwidth]{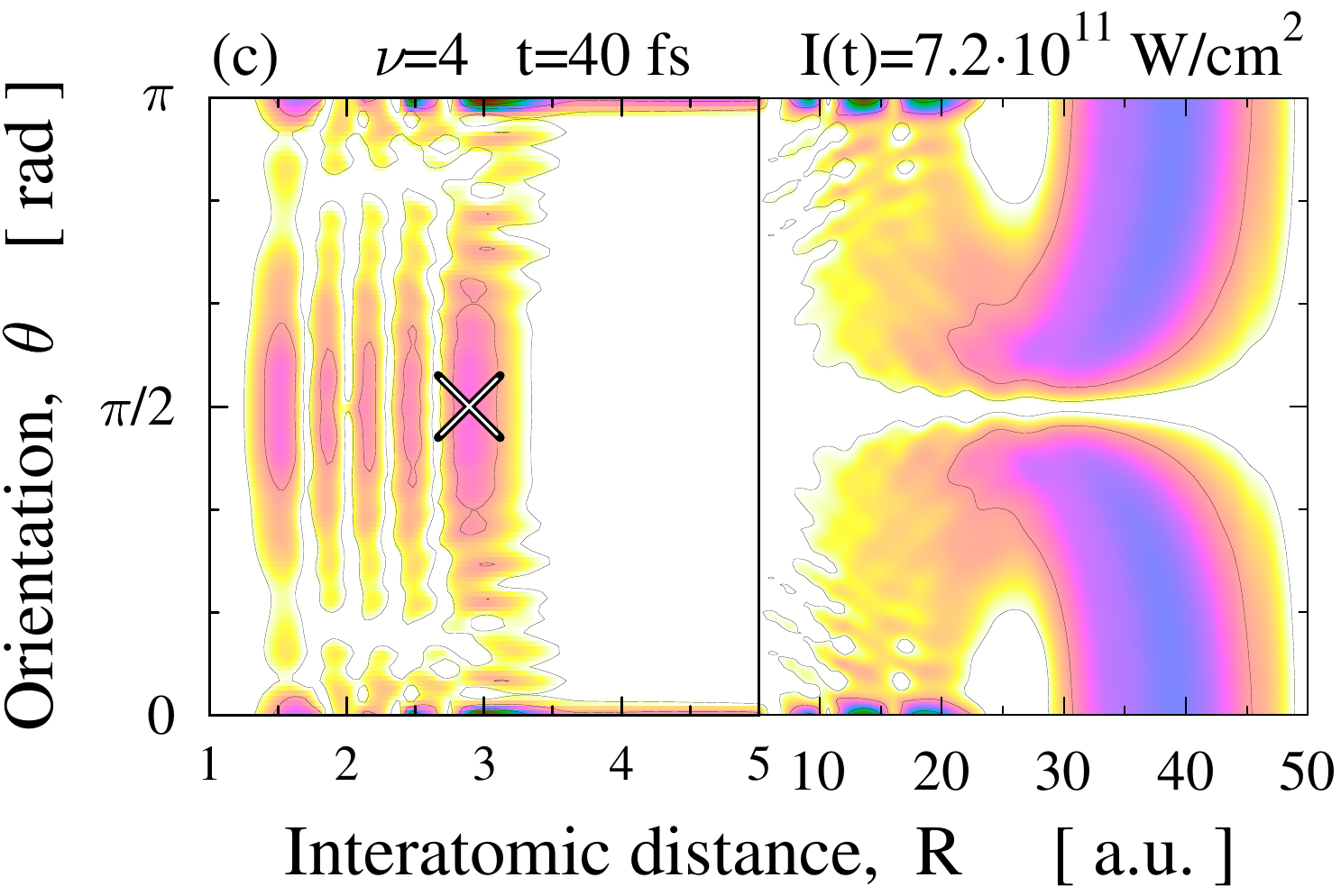}
\par\end{center}%
\end{minipage}\hfill{}%
\begin{minipage}[c]{0.07\textwidth}%
\begin{center}
\includegraphics[width=0.45\textwidth]{wf_0}
\par\end{center}%
\end{minipage}
\par\end{centering}

\caption{\label{fig:4}Snapshots from the real-time evolution of the nuclear
wave packet density. I=$1\times10^{14}W/cm^{2}$ intensity and $\nu=4$
vibrational state are applied. The cross denotes the position of the
LICI. }
\end{figure*}
It is worth mentioning that even at high intensities the sign of the
interference in the angular distribution does not always appear. For
example for the level of $\nu=0$ due to bond softening the dissociation
takes place near parallel to the polarization of the field, and the
dissociation probability changes so rapidly as the function of the
angle that the small modulation caused by the rotational nodes is
not visible in the angular distribution (Fig. \ref{fig:2}, I=$1\times10^{14}W/cm^{2}$).
On the other hand, for $\nu=4$ we obtain very rich interference patterns
in the nuclear wave packet density (Fig. \ref{fig:4}) but owing to
integration, the fingerprints of this interference disappears from
the angular distribution (Fig. \ref{fig:2}).

\begin{figure}[t]
\begin{centering}
\includegraphics[width=0.37\textwidth]{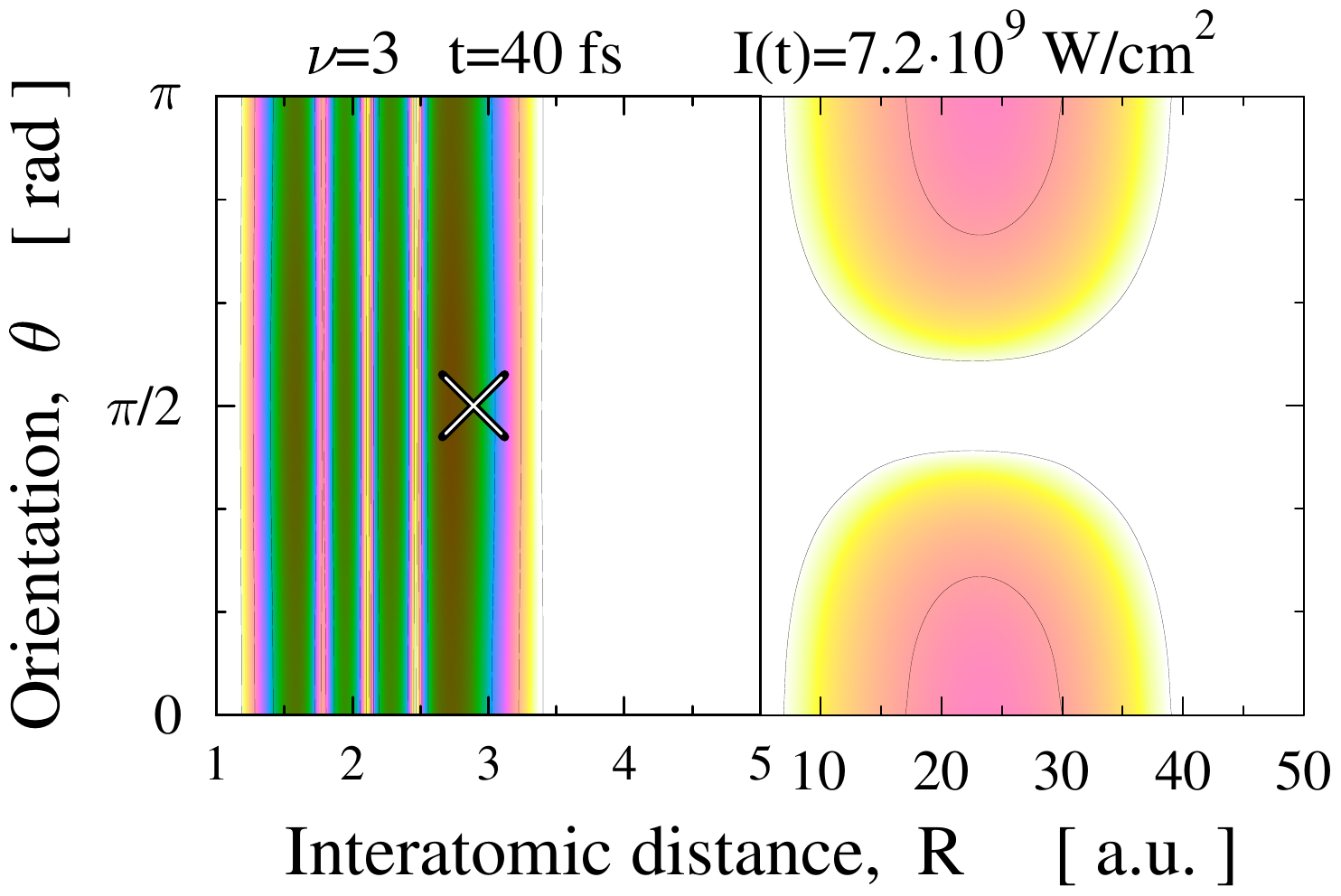}~~\includegraphics[width=0.04\textwidth]{wf_0}
\par\end{centering}

\caption{\label{fig:5}Snapshot from the real-time evolution of the nuclear
wave packet density. I=$1\times10^{12}W/cm^{2}$ intensity and $\nu=3$
vibrational state are applied. The cross denotes the position of the
LICI. }
\end{figure}
We note here, that the situation is quite different for small and
moderate intensity values as such I=$1\times10^{12}W/cm^{2}$. For
any given vibrational state, there is no interference pattern appeared
in the nuclear density function (see e.g. Fig. \ref{fig:5} for $\nu=3$).
The applied field is not strong enough to build up notable rotational
nodes. 

In summary, the results obtained clearly confirm that additional structures
or modulations appear in the angular distributions strongly related
to the interference properties of the nuclear wave packet. 

Additionally, one dimensional calculations (1d) for the angular distribution
have also been performed. We froze the rotational degree of freedom
and accordingly the LICI was not considered. With such conditions,
the molecule's initial orientation was not changing during the dissociation
process and the ``effective field strength'' was the projection
of the real field to the axis of the molecule: $\epsilon_{0}^{eff}=\epsilon_{0}\cdot\cos\theta$
$\left(I_{0}^{eff}=I_{0}\cdot\cos^{2}\theta\right)$. In such cases,
we have never obtained modulations on the angular distribution curves
similar to the ones observed in the full calculations (Fig. \ref{fig:2}).
(The scale on Fig. \ref{fig:2} was chosen so that the dissociation
rate of 1 represents the total dissociation in a given direction.
Larger values in the full calculation mean that some parts of the
dissociating particles were rotated by the field to this direction
from some different initial ones.)

The main issue has been discussed in this letter is related to the
rotation. By means of two dimensional quantum dynamical calculations,
we have demonstrated that the additional rotational degree of freedom
in the description of the $\mathrm{D}_{2}^{+}$ photodissociation
promises a wealth of novel and observable quantum phenomena. After
intense laser-matter interaction, a measurable quantum interference
effect occurs in the molecular system due to the very strong nonadiabatic
coupling of the electronic, rotational and vibrational motions.
\begin{acknowledgments}
The authors acknowledge financial support by the Deutsche Forschungsgemeinschaft
(Project ID CE10/50-1). Á. V. acknowledges the computational resources
provided by the John-von-Neumann Institute, Research Centre Juelich
(Project ID ehu01), the OTKA Grant no. NN103251 and the TÁMOP 4.2.2.C-11/1/KONV-2012-0001.
N. M. thanks for the ISF grant 298/11.\end{acknowledgments}

\end{document}